\documentstyle[preprint,aps,floats,psfig,epsf]{revtex}
\catcode`@=11
\def\references{%
\ifpreprintsty
\bigskip\bigskip
\hbox to\hsize{\hss\large \refname\hss}%
\else
\vskip24pt
\hrule width\hsize\relax
\vskip 1.6cm
\fi
\list{\@biblabel{\arabic{enumiv}}}%
{\labelwidth\WidestRefLabelThusFar  \labelsep4pt %
\leftmargin\labelwidth %
\advance\leftmargin\labelsep %
\ifdim\baselinestretch pt>1 pt %
\parsep  4pt\relax %
\else %
\parsep  0pt\relax %
\fi
\itemsep\parsep %
\usecounter{enumiv}%
\let\p@enumiv\@empty
\def\theenumiv{\arabic{enumiv}}%
}%
\let\newblock\relax %
\sloppy\clubpenalty4000\widowpenalty4000
\sfcode`\.=1000\relax
\ifpreprintsty\else\small\fi
}
\catcode`@=12

\def\lsim{\mathrel{\raise.3ex\hbox{$<$\kern-.75em\lower1ex\hbox{$\sim$}}}}
\def\gsim{\mathrel{\raise.3ex\hbox{$>$\kern-.75em\lower1ex\hbox{$\sim$}}}}
\begin{document}
\tightenlines

\hfill\vtop{
\hbox{MADPH-01-1223}
\hbox{AMES-HET-01-05}
\hbox{hep-ph/0104166}
\hbox{}}

\vspace*{.25in}
\begin{center}
{\large\bf Piecing the Solar Neutrino Puzzle Together at SNO}\\[10mm]
V. Barger$^1$, D. Marfatia$^1$ and K. Whisnant$^2$\\[5mm]
\it
$^1$Department of Physics, University of Wisconsin,
Madison, WI 53706, USA\\
$^2$Department of Physics and Astronomy, Iowa State University,
Ames, IA 50011, USA

\end{center}
\thispagestyle{empty}

\begin{abstract}

\vspace*{-.35in}

\noindent

We perform an oscillation parameter-independent 
analysis of solar neutrino flux measurements 
from which we predict the charged-current rate at SNO relative to
Standard Solar Model to be \mbox{$R_{\rm SNO}^{CC} =
0.35^{+0.16}_{-0.09} (0.46^{+0.13}_{-0.11})$} for oscillations to
active (sterile) neutrinos.
By alternately considering the $^8$B flux normalization fixed and free, we find that 
the flux measured by Super-Kamiokande (SK) not being a result of oscillations
is strongly disfavored for oscillations to active neutrinos. 
 SNO will determine the best-fit value of the $^8$B flux normalization $\beta$
(equal to the neutral-current rate), 
without recourse to 
neutral-current measurements, from the derived relation 
$\beta=R_{\rm SNO}^{NC}=5.85 R_{\rm SK} - 4.85 R_{\rm SNO}^{CC}$.
Using a simple parameterization of the fraction of high,
intermediate, and low energy solar neutrinos starting above resonance, we 
reproduce the results of global analyses 
to good accuracy; we find that the LMA solution with a normal mass 
hierarchy is clearly favored.  
With $\beta$ free, our analysis for oscillations to
active neutrinos gives $R_{\rm SNO}^{NC}=\beta=1.34 \pm 0.34$, which corresponds to 
$R_{\rm SNO}^{CC}=0.28\pm 0.07$.

\end{abstract}

\newpage

\section{Introduction}

Solar neutrino experiments measure an energy-dependent flux suppression
~\cite{homestake,sage,gallex,gno,sk} relative to the Standard Solar
Model (SSM)~\cite{SSM}.  This situation has existed for
thirty years and every new solar neutrino experiment has confirmed the
flux-deficit. The best motivated explanation of this solar neutrino puzzle
is that neutrinos are massive and undergo oscillations.
The energy dependence of the flux
suppression singles out very specific regions in the space of parameters
that govern the frequency and amplitude of neutrino
oscillations. 
 The solar-neutrino flux deficit can be accounted for by
oscillations of electron neutrinos to mu and/or tau neutrinos or to
sterile neutrinos that do not interact weakly. In the two-neutrino
oscillation framework, oscillations into sterile neutrinos are excluded at
the 95\% C. L.~\cite{s} by a comparison of the day and night spectra at
Super-Kamiokande (Super--K) and the results of a global flux analysis; hence 
the favored explanation is
\mbox{$\nu_e \rightarrow \nu_{\mu}/ \nu_{\tau}$}
oscillations. For such oscillations, there are three regions 
(LMA, SMA and LOW) in the mass-squared difference $\delta m^2$ 
 and vacuum amplitude $\sin^22\theta$ parameters{\footnote{From a flux-independent 
analysis~\cite{analysis}, the best-fit
values of $(\delta m^2\  {\rm{eV}}^2,\sin^22\theta)$ for the three solutions are
$(4.2\times 10^{-5},0.66)$ (LMA), $(5.2\times 10^{-6},2.2\times 10^{-3})$ (SMA) and
$(7.6\times 10^{-8},0.97)$ (LOW).}}. These solutions involve effects 
from coherent $\nu_e$ scattering from matter~\cite{bppw} in the Earth and the Sun.
Of these regions, the SMA
region is disfavored at the 95\% C. L. because of the observed flat energy spectrum
and imperceptible day/night effect at Super--K~\cite{s}. For this reason
 we drop further consideration of the SMA region.
 A fourth region, the VAC or Just-So solution, which is largely independent of
matter effects, is also excluded at the 95\% C. L. by the same
considerations that disfavor the SMA solution. Thus, we are left with
the LMA and LOW regions, both of which have large mixing. 
With a large measure of certainty the
 KamLAND~\cite{kamland} experiment will exclude or confirm the LMA region as a
solution to the solar neutrino anomaly~\cite{kam,comp}. The SNO
experiment~\cite{sno} will be crucial in validating the above
 conclusions from the Super--K experiment~\cite{oursno,smirnov}.

In this Letter we first perform a simple neutrino oscillation-independent analysis 
(with SSM fluxes) of the
solar neutrino data using the total rates at the
$^{37}$Cl~\cite{homestake} and $^{71}$Ga~\cite{sage,gallex,gno}
experiments and Super--K, following the procedure proposed in 
Ref.~\cite{bpw91}.  We make predictions for the charged-current
(CC) rate at the SNO experiment. Allowing the $^8$B flux normalization to be free, we 
 derive a relation for the neutral-current (NC) rate at SNO in terms of the CC rate at SNO in a model-independent way. 
Our analysis is suited to the LMA and LOW
solutions for which the oscillations in matter are mainly adiabatic. 
We find the
relative flux suppression of the high, intermediate, and low energy
solar neutrinos, compared to the SSM, and how these suppressions depend
on the normalization of the solar $^8$B neutrino flux.
 We then apply our analysis to the LMA
and LOW solutions and approximately reproduce the results of
more comprehensive fits. Finally, we predict the CC and NC rates at SNO for the
$^8$B flux normalization obtained by imposing adiabatic constraints. 

\section{Model-Independent Analysis}
\label{sec:analysis}

Following the procedure of Ref.~\cite{bpw91}, we divide the solar
neutrino spectra into three parts: high energy (consisting of $^8$B and
$hep$ neutrinos), intermediate energy ($^7$Be, $pep$, $^{15}$O, and
$^{13}$N), and low energy ($pp$). For each class of solar neutrino
experiment the fractional contribution without oscillations to the
expected rate from each part of the spectrum can be calculated in the
SSM (see Table~\ref{tab:spectrum}). If $R$ is the measured rate divided
by the SSM prediction for a given experiment, then with oscillations,
\begin{eqnarray}
R_{\rm Cl} &=& 0.764 P_H + 0.236 P_I \,,
\label{eq:RCl}\\
R_{\rm Ga} &=& 0.096 P_H + 0.359 P_I + 0.545 P_L \,,
\label{eq:RGa}
\end{eqnarray}
where $P_H$, $P_I$, and $P_L$ are the average survival probabilities
for the high, intermediate, and low energy solar neutrinos. In
Eqs.~(\ref{eq:RCl}) and (\ref{eq:RGa}) we assume that each probability
remains the same from experiment to experiment, which is justified since
 the differential event rate without oscillations for each part
of the spectrum has approximately the same shape for all experiments~\cite{bpw91}. 

The SNO experiment detects neutrinos with energy above 5 MeV
 primarily via the reaction 
$\nu_e + d \rightarrow p+p+e^-$; the predicted CC rate is
\begin{equation}
R_{\rm SNO}^{CC} = P_H \,.
\label{eq:RSNO}
\end{equation}
\begin{table}[b]
\caption[]{\label{tab:spectrum}
Fractional contribution of the high, intermediate and low energy
neutrinos to the $^{37}$Cl, $^{71}$Ga and Super--K signals without
oscillations. The last column gives the normalization uncertainty for
each part of the spectrum.}
\begin{tabular}{l|l|ccc|c}
&&&& Super--K & Normalization\\
& & $^{37}$Cl & $^{71}$Ga & SNO & Uncertainty\\
\hline
High
& $^8$B, $hep$ & 0.764 & 0.096 & 1.000 & 18.0\%\\
Intermediate
& $^7$Be, $pep$, $^{15}$O, $^{13}$N & 0.236 & 0.359 & 0.000 & 11.6\%\\
Low
& $pp$ & 0.000 & 0.545 & 0.000 & 1.0\%
\end{tabular}
\end{table}
$\!\!$The $R$ formula for Super--K is different from SNO, even
though both experiments are sensitive to only the high energy neutrinos, 
because Super--K detects
active neutrinos by elastic scattering, $\nu_x + e^- \rightarrow \nu_x' + e^-$ 
($\nu_x$ denotes any of the active flavors), 
and that has neutral-current contributions. 
Using the NC/CC cross section ratio of 0.171 (for $x_W =
0.225$), we have
\begin{eqnarray}
R_{\rm SK} &=& 0.829 P_H + 0.171 \ \ \ \, {\rm~(active)} \,,
\label{eq:RSKa}\\
           &=& P_H \ \ \ \ \ \ \ \ \ \ \ \ \ \ \ \ \ \ \ \ \ {\rm~(sterile)} \,.
\label{eq:RSKs}
\end{eqnarray}

The solar neutrino data are summarized in Table~\ref{tab:data}.
Since there are three probability unknowns and three data points, there is an 
unique solution for the $P_j$. The
Super--K rate depends only on the high energy neutrinos, so $P_H$ is
directly determined by $R_{\rm SK}$. Then since $R_{\rm Cl}$ depends
only on $P_I$ and $P_H$, the value of $P_I$ may be determined from $R_{\rm Cl}$.
Finally, $P_L$ may be determined from $R_{\rm Ga}$.
 The best-fit values to the data in Table~\ref{tab:data} are
\begin{eqnarray}
P_H &=& 0.347 \,,\  P_I = 0.303 \,,\ \ \  P_L = 0.811 \ \ \, {\rm~(active)} \,.
\label{eq:active}\\
P_H &=& 0.459 \,,\  P_I = -0.058 \,,\  P_L = 1.029 \ \ {\rm~(sterile)} \,,
\label{eq:sterile}
\end{eqnarray}
The sterile solution lies somewhat outside of, although close to, the
physical region.
\begin{table}[t]
\begin{center}
\begin{eqnarray}
\begin{array}{lc|cc}
\rm{Experiment} & & &\rm{data/SSM}\\
\hline
^{37}\rm{Cl}  & & &0.337 \pm 0.030\\
^{71}\rm{Ga} & & &0.584 \pm 0.039\\
{\rm{Super\!-\!K}}  & & & 0.459 \pm 0.017 \nonumber
\end{array}
\end{eqnarray}
\caption[]{\label{tab:data}
Solar neutrino data~\cite{homestake,sage,gallex,gno,sk} 
expressed as the ratio $R =$~data/SSM, including the
experimental uncertainties. The $^{71}$Ga number combines the results
of the GALLEX, SAGE, and GNO experiments.}
\end{center}
\end{table}

To determine the allowed regions in probability space, we include
the experimental uncertainties in the measured values 
 and the theoretical uncertainties of the high,
intermediate, and low energy neutrino fluxes in the SSM (see
Table~\ref{tab:spectrum}). We use the following expression for $\chi^2$:
\begin{equation}
\chi^2 = \sum_i {(R_i - R_i^{th})^2 \over
\delta R_i^2 + a_i^2 P_H^2 \delta_H^2+ b_i^2 P_I^2 \delta_I^2
+ c_i^2 P_L^2 \delta_L^2} \,.
\label{eq:chi2}
\end{equation}
Here $i$ runs over the three types of experiments ($^{37}$Cl,
$^{71}$Ga, and Super--K), $R_i$ and $\delta R_i$ are the
central values and uncertainties of data/SSM, $\delta_j$ is the
normalization uncertainty of that part of the solar spectrum
($j = H, I, L$), and the theoretical prediction for $R_i$ is
\begin{equation}
R_i^{th} = a_i P_H + b_i P_I + c_i P_L + d_i \,. 
\label{eq:Ri}
\end{equation}
The coefficients in Eq.~(\ref{eq:Ri}) are given in
Eqs.~(\ref{eq:RCl}), (\ref{eq:RGa}), (\ref{eq:RSKa}), and
(\ref{eq:RSKs}), and the normalization uncertainties are given in
Table~\ref{tab:spectrum}.

\begin{figure}[t]
\mbox{\psfig{file=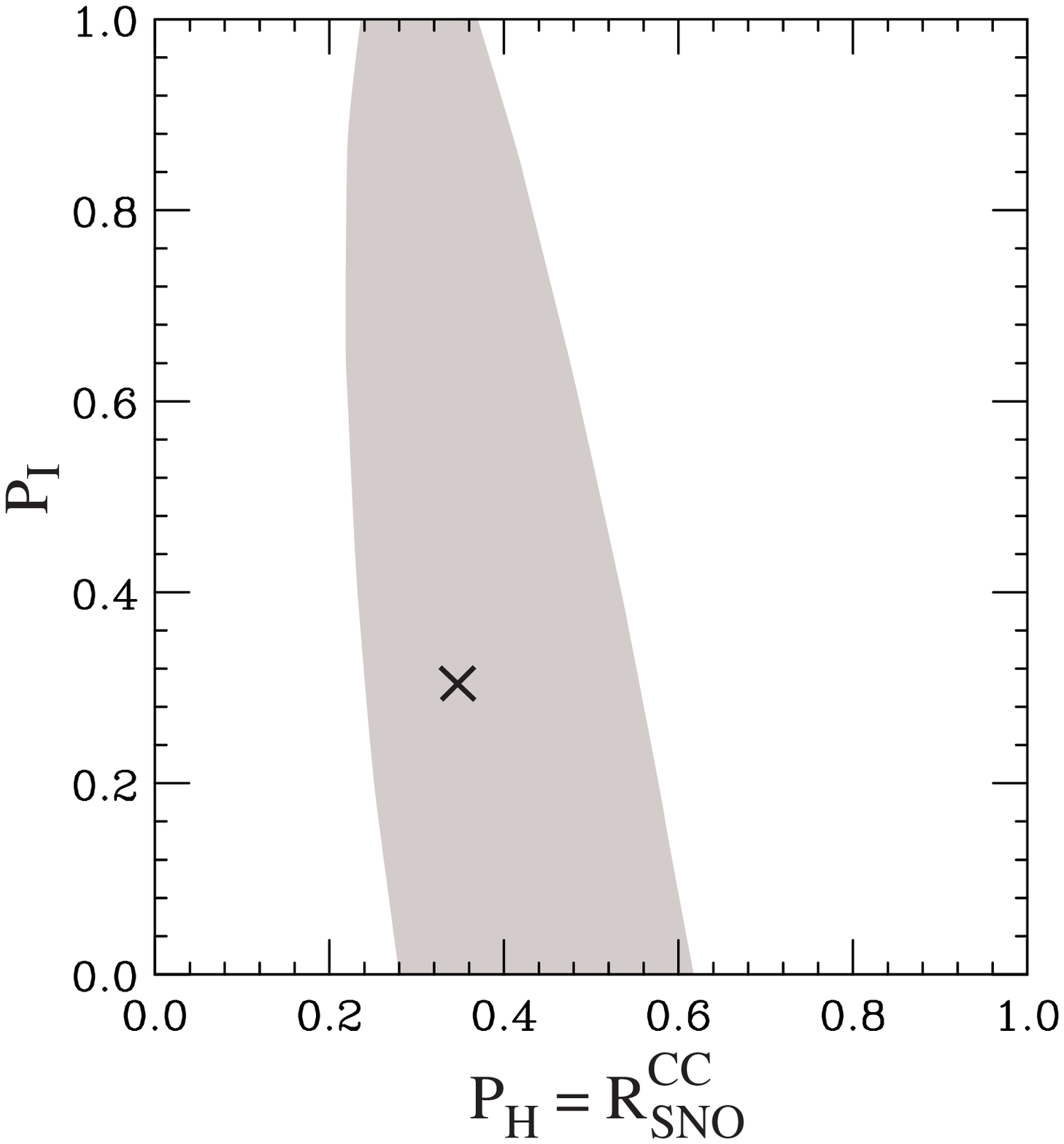,width=5.5cm,height=5.5cm}
\psfig{file=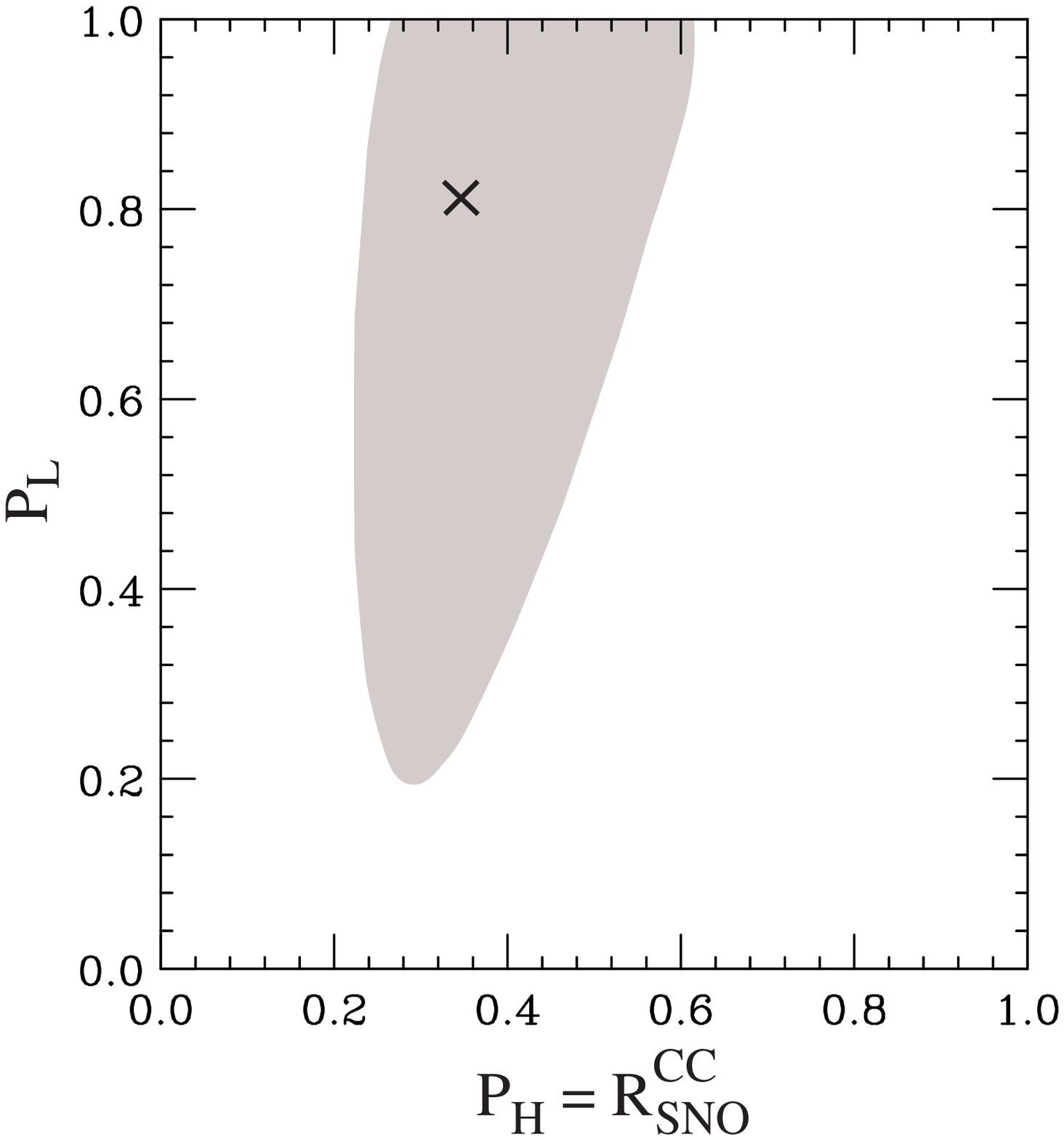,width=5.5cm,height=5.5cm}
\psfig{file=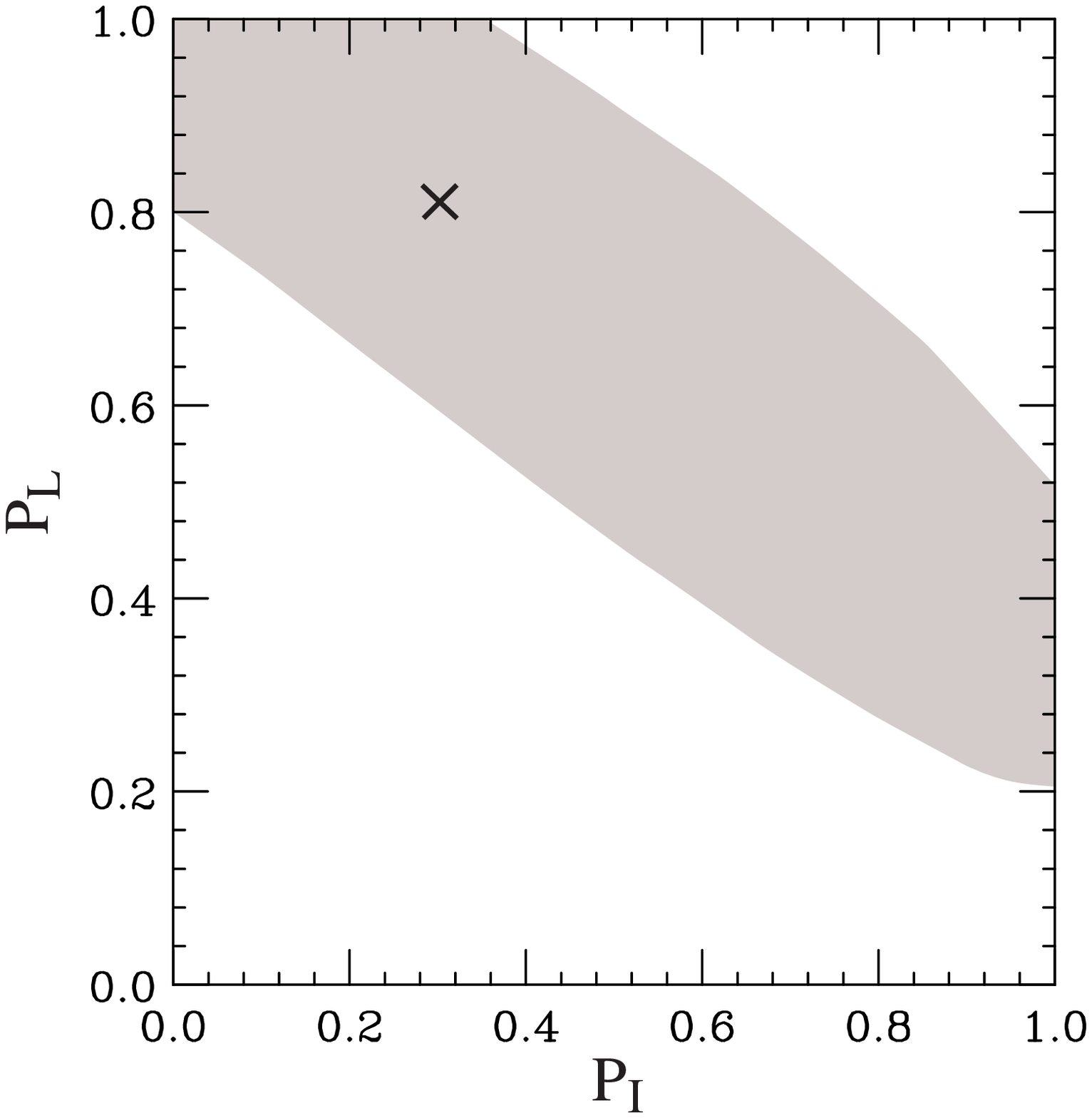,width=5.5cm,height=5.5cm}}
\medskip
\caption{Best-fit values and 95.4\%~C.L. allowed regions for $P_I$
versus $P_H$, $P_L$ versus $P_H$, and $P_L$ versus $P_I$,
assuming oscillations to active neutrinos and SSM fluxes ($\beta=1$).
}
\label{fig:active}
\end{figure}

\begin{figure}[tb]
\mbox{\psfig{file=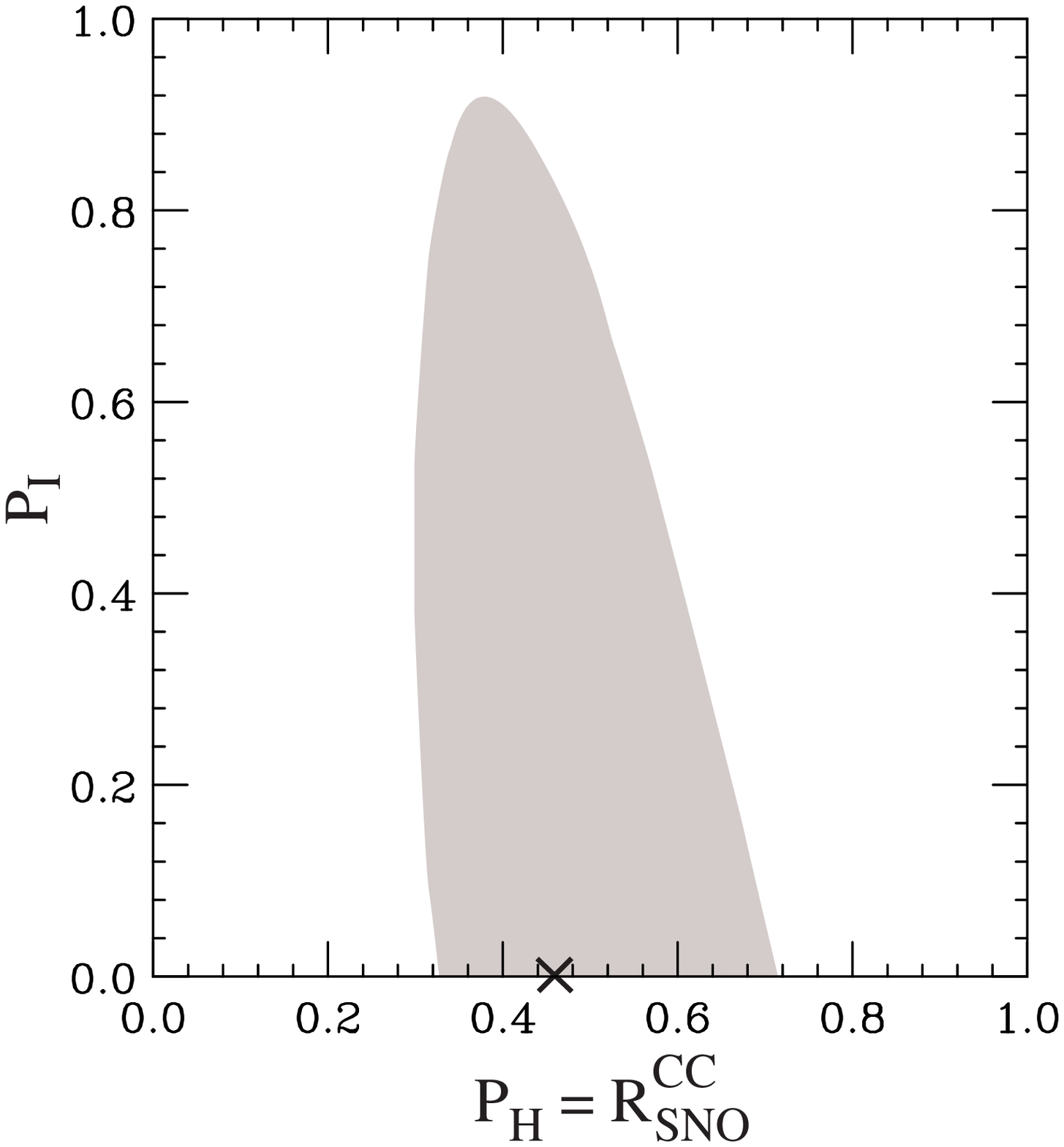,width=5.5cm,height=5.5cm}
\psfig{file=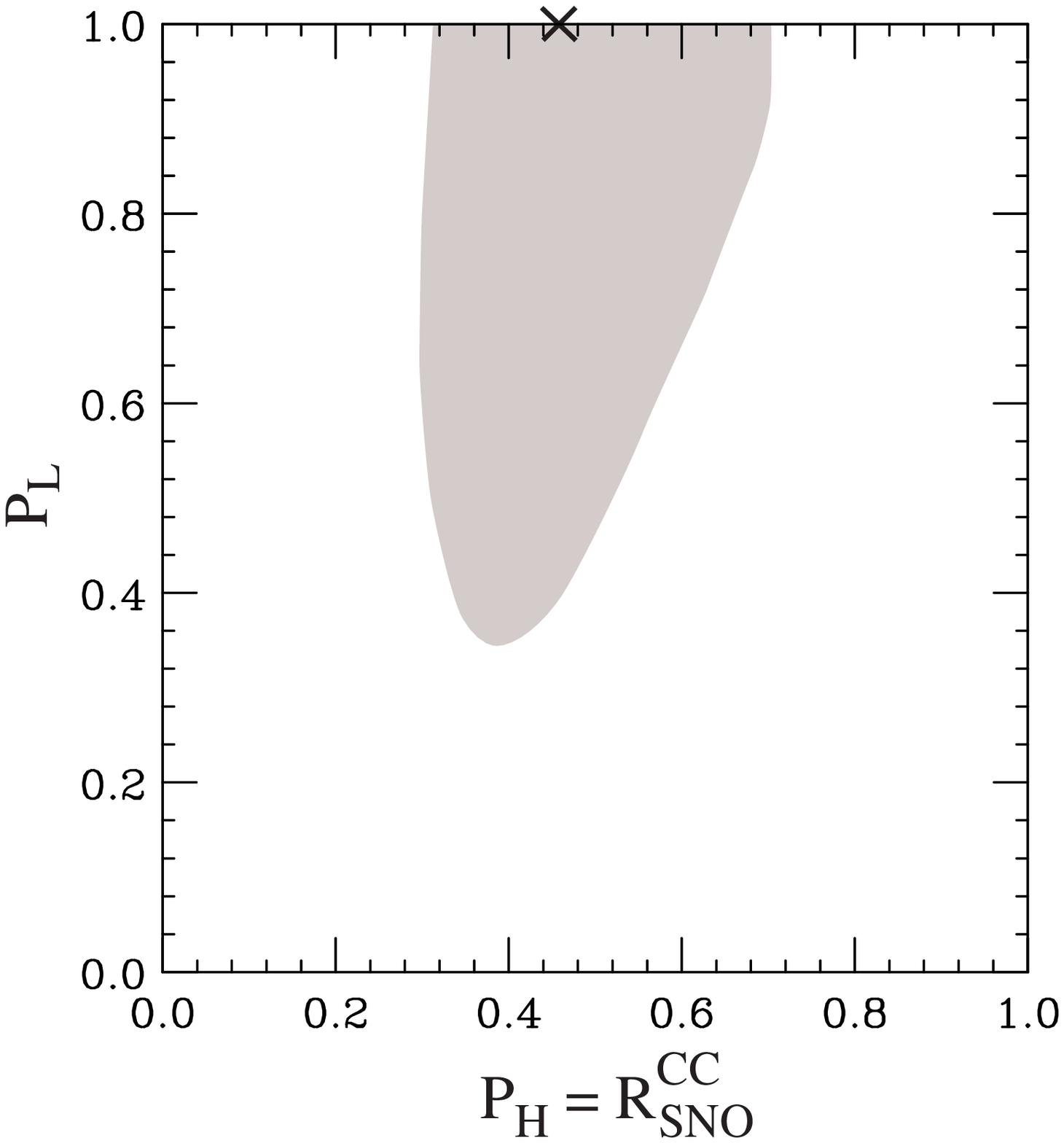,width=5.5cm,height=5.5cm}
\psfig{file=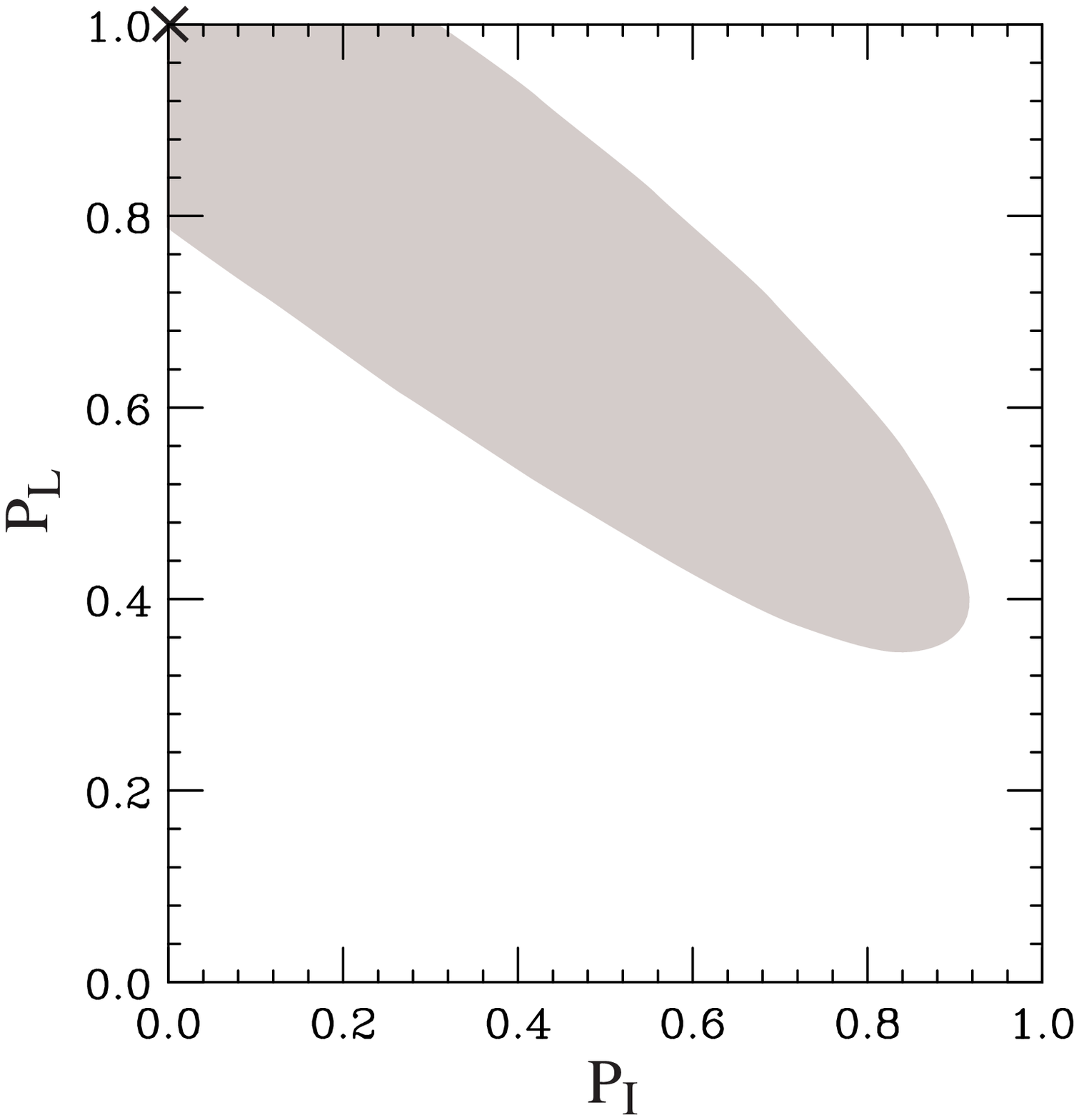,width=5.5cm,height=5.5cm}}
\medskip
\caption{Same as Fig.~\ref{fig:active} for oscillations to sterile
neutrinos.
}
\label{fig:sterile}
\end{figure}

The best-fit values and 95.4\%~C.L. ($2\sigma$) allowed regions ($\Delta \chi^2< 8.02$), 
for each of the
two-dimensional subspaces of the three-dimensional probability space are shown in
Figs.~\ref{fig:active} and \ref{fig:sterile} for oscillations to active
and to sterile neutrinos, respectively. Note that the uncertainty in $P_I$ spans
 the entire physically allowed region for active neutrino oscillations. 
The Borexino~\cite{borexino} and KamLAND~\cite{kamland} experiments will 
target $^7$Be solar neutrinos and should be able to narrow this range significantly. 
 
SNO observes only high energy
neutrinos. Our analysis with the SSM fluxes yields the simple prediction
\begin{eqnarray}
R_{\rm SNO}^{CC} = P_H &=& 0.35 \, {\rm~(active)} \,
\label{eq:SNOa}\\
                  &=& 0.46 \, {\rm~(sterile)} \,
\label{eq:SNOs}
\end{eqnarray}
with a possible range
\begin{eqnarray}
0.23 &\le& R_{\rm SNO}^{CC} \le 0.62 \, {\rm~(active)} \,
\label{eq:SNOrangea}\\
0.30 &\le& R_{\rm SNO}^{CC} \le 0.71 \, {\rm~(sterile)} \,
\label{eq:SNOranges}
\end{eqnarray}
at 95.4\%~C.L.. Because the ranges in Eqs.~(\ref{eq:SNOrangea}) and
(\ref{eq:SNOranges}) are partially non-overlapping, a 
differentiation between the active and sterile scenarios from 
CC data alone could be possible.

\section{Uncertainty of the $^8$B flux}

\subsection{$^8$B flux normalization fixed by Super--K}

One approach to the uncertainty of the $^8$B flux is to assume that
the $^8$B neutrinos are not suppressed at all by oscillations (or other
particle physics mechanisms), and hence that the Super--K experiment is
providing a direct measurement of the $^8$B flux. Then the implications
of the $^{37}$Cl and $^{71}$Ga data on $P_I$ and $P_L$ can be examined.

This scenario can be modeled by fixing $P_H= 0.459$, the
value of data/SSM measured by Super--K. Then the $\chi^2$ for the
remaining two types of experiments can be evaluated by using
Eq.~(\ref{eq:chi2}), with $\delta_H$ set equal to the fractional
uncertainty in the Super--K measurement, i.e., $\delta_H = 0.017/0.459 =
0.037$, and summing over $i = I,L$. There are two degrees of
freedom, $P_I$ and $P_L$, and two data points. The unique
solution, with $\chi^2 = 0$, is
\begin{equation}
P_I = -0.058, \, P_L = 1.029 \,.
\end{equation}
While this best-fit solution lies outside the physical region, a
solution with $P_I = 0$ and $P_L = 1$ is nearly as good, giving a $\chi^2$ of
0.2. At 95.4\%~C.L. we find that $P_I < 0.30$ and $P_L > 0.74$. Hence for
this scenario to be a good description of the data, the
intermediate energy neutrinos are strongly suppressed, while the
low energy neutrinos are not greatly suppressed. Only a vacuum solution with 
$\delta m^2 \sim 6 \times 10^{-12}\  {\rm{eV}}^2$ and large mixing can give such 
probabilities~\cite{justso}; however, the global analysis of Ref.~\cite{analysis} 
shows that this solution is acceptable only at 
\mbox{99\% C. L.} for active neutrino oscillations (90\% C. L. for oscillations to 
sterile neutrinos). 
If none of the low or intermediate energy neutrinos
are suppressed, the $\chi^2$ is 91.2. Thus the scenario in which the
$^8$B flux is not suppressed by oscillations 
and is being directly measured by Super--K 
is highly disfavored for oscillations to active neutrinos.

\subsection{Varying the $^8$B flux normalization}

The uncertainty in the $^8$B flux is relatively large (18\% at the
$1\sigma$ level), and it is possible that the SSM does not give
a good estimate of it. Rather than just include the $^8$B flux 
uncertainty in the calculation
of $\chi^2$, we can allow the $^8$B flux normalization to be a
free parameter in the fit. We define $\beta$ to be the
$^8$B flux relative to the SSM. For oscillations to active neutrinos, the neutral current
rate (NC/SSM) at SNO is
\begin{equation}
R_{\rm SNO}^{NC} = \beta \ \ \ \ \ \, {\rm~(active)} \,.
\end{equation}
$\chi^2$ for oscillations to active
neutrinos is determined by using Eq.~(\ref{eq:chi2}) with the changes
$\delta_H \to 0$ (the uncertainty in the $^8$B flux is to be determined
by the fit), $P_H \to \beta P_H$ in Eqs.~(\ref{eq:RCl}) and
(\ref{eq:RGa}), and $R_{\rm SK} \to \beta (0.829 P_H + 0.171)$ in
Eq.~(\ref{eq:RSKa}).

Since there are now four parameters and only three data points, there is
no longer a unique solution, but a family of solutions parameterized by
$\beta$. The values of $P_H$, $P_I$, and $P_L$ that exactly reproduce
the data are plotted versus $\beta$ in Fig.~\ref{fig:beta}. The figure
shows that for no value of $\beta$ are the $P_j$ all equal (although two
of the three could be equal), so that there must be an energy-dependent
suppression compared to the SSM. Values of $\beta \lsim 1$
(i.e., initial $^8$B flux less than the SSM) imply a greater
suppression of the intermediate energy oscillation probability while
for $\beta \gsim 1$ the high energy probability is more
suppressed. Figure~\ref{fig:beta}
represents the possible exact solutions of the solar neutrino puzzle
against which particular models can be compared. The only caveat is that
our analysis accounts only for the average rates for each part of the
neutrino spectrum; other measurements, such as
 energy dependence within one part of the spectrum or the
day/night asymmetry, may provide further constraints.

\begin{figure}[t]
\mbox{\psfig{file=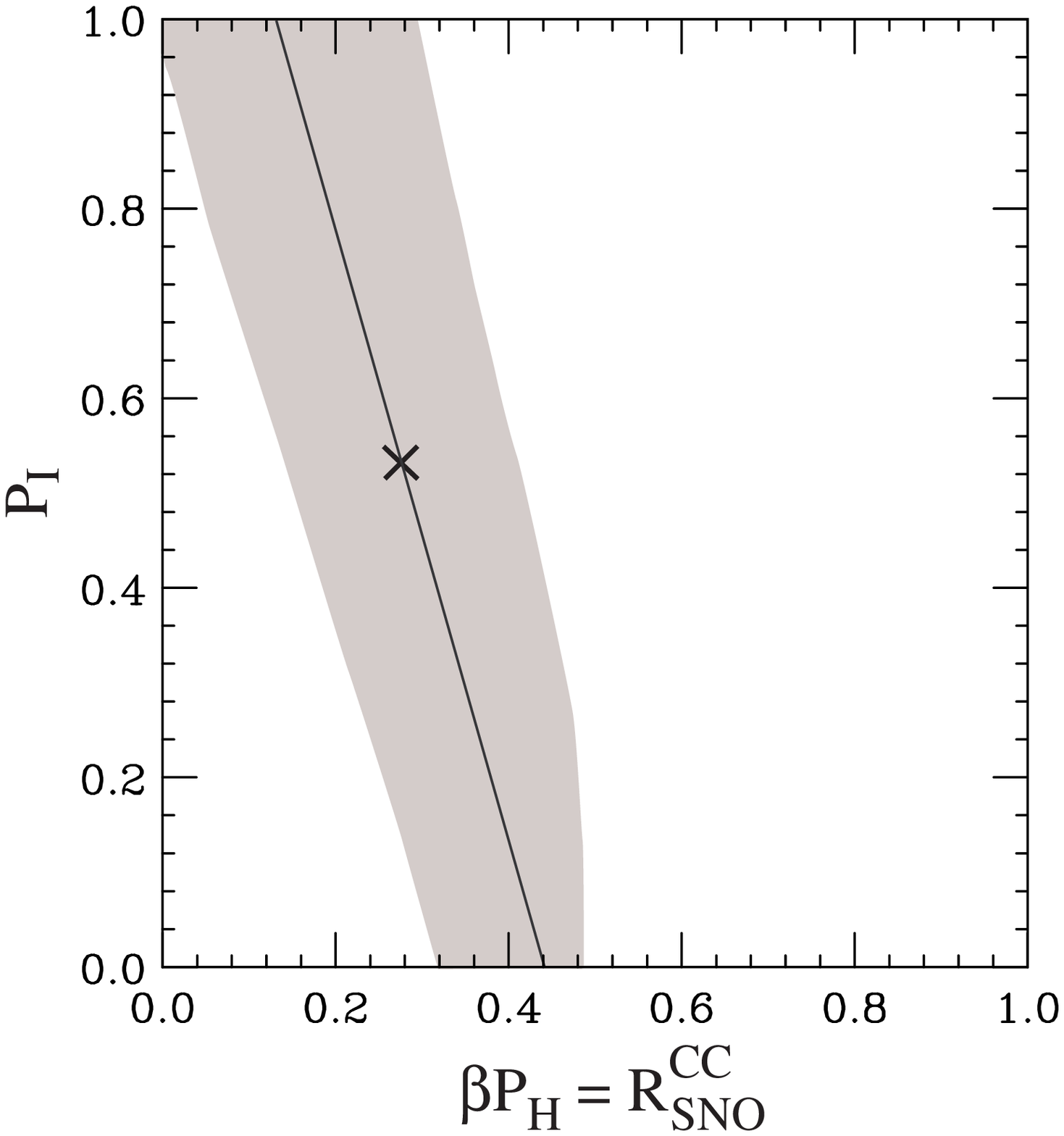,width=8cm,height=8.cm}
\psfig{file=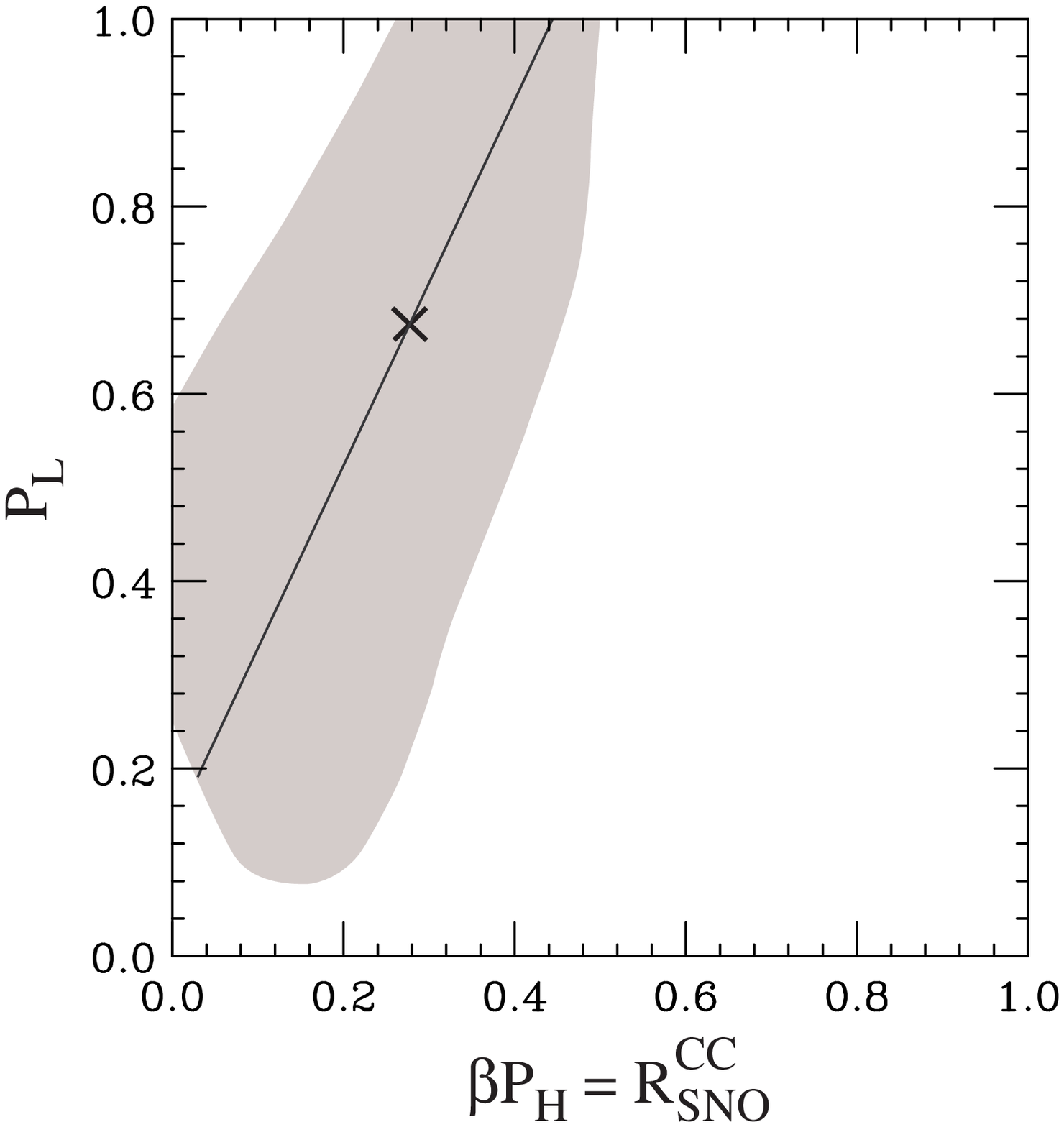,width=8cm,height=8.cm}}
\medskip
\mbox{\psfig{file=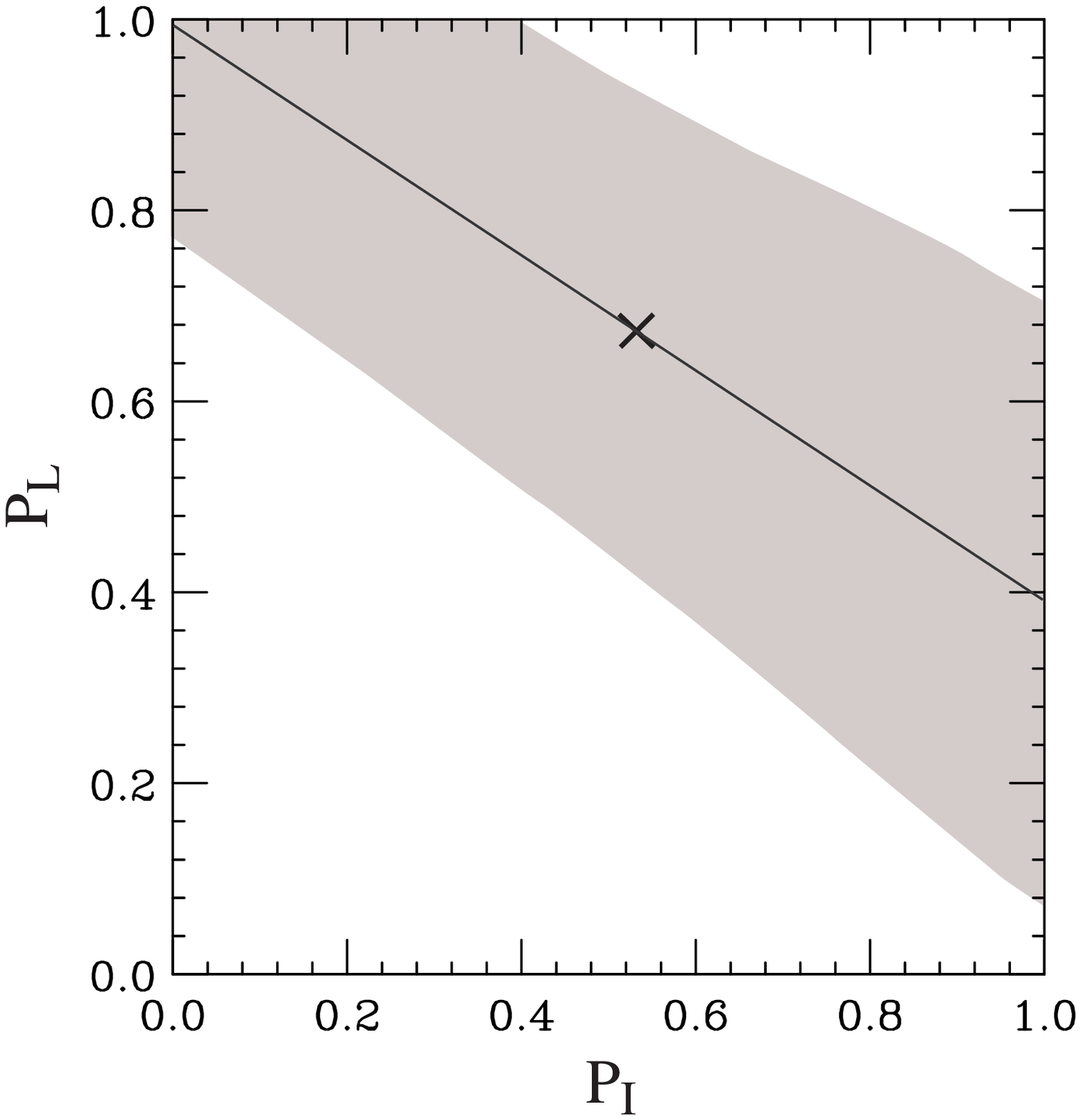,width=8cm,height=8.cm}
\psfig{file=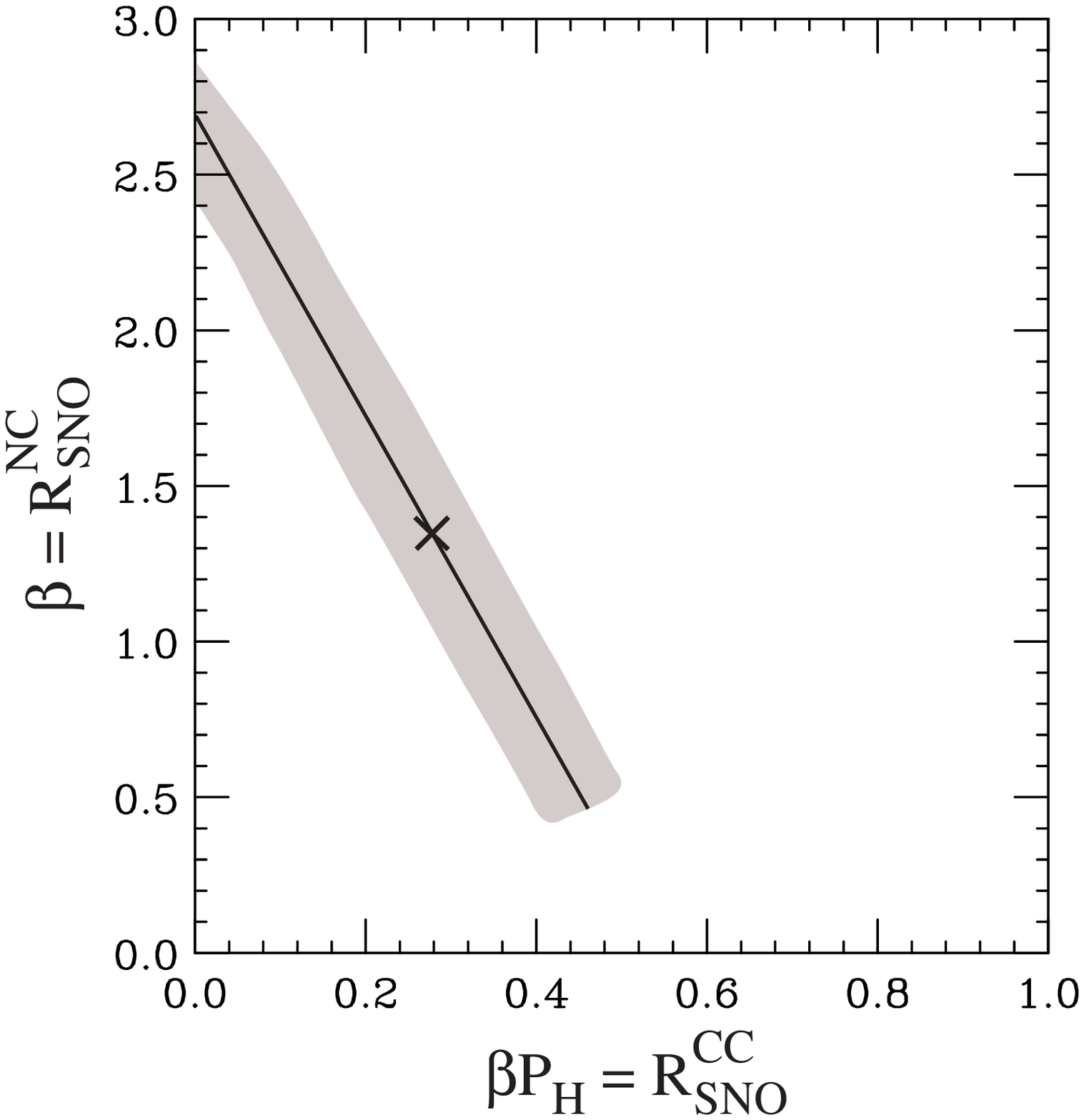,width=8cm,height=8.cm}}
\caption{The 95.4\%~C.L. allowed regions from an analysis with 
 free $^8$B flux normalization $\beta$, for $P_I$
versus $\beta P_H$, $P_L$ versus $\beta P_H$, $P_L$ versus $P_I$ and
 $\beta$ versus $\beta P_H$, assuming oscillations to active neutrinos. The line represents
solutions with $\chi^2=0$. The cross marks the point selected once adiabatic constraints are
imposed.
}
\label{fig:fluxnorm}
\end{figure}

The family of exact solutions give a prediction for SNO that depends only on
$\beta$ and $R_{\rm SK}$,
\begin{eqnarray}
R_{\rm SNO}^{CC} = \beta P_H &=& 1.21 R_{\rm SK}- 0.21 \beta \\
                        &=&  0.55 - 0.21 \beta\ \ \ \ \ \ \ \  (R_{\rm SK}=0.459)\,.
\label{eq:SNObeta}
\end{eqnarray}
Thus the SNO CC measurement will select a particular best-fit value of
the $^8$B flux normalization and the NC rate, 
\begin{eqnarray}
R_{\rm SNO}^{NC}=\beta&=&5.85 R_{\rm SK}- 4.85 R_{\rm SNO}^{CC}\\
                      &=&2.69 - 4.85 R_{\rm SNO}^{CC}\ \ \ \ \ \ \ \  (R_{\rm SK}=0.459)\,.
\end{eqnarray}
Performing a $^8$B flux-independent analysis by varying $P_H$, $P_I$, $P_L$ and $\beta$,
the predicted CC rate at SNO has the upper bound
\begin{equation}
 R_{\rm SNO}^{CC} \leq 0.5\ \ \ \ \ \ \ (^8{\rm{B\ flux\  free}})
\end{equation}
at 95.4\%~C.L.. The large allowed range of $R_{\rm SNO}^{CC}$ represents the variation of 
$\beta$ from 0.4 to 2.8 (see the plot of $\beta$ versus $\beta P_H$ in 
Fig.~\ref{fig:fluxnorm}).
The best-fit line with $\chi^2=0$ (since there are four parameters and three constraints), 
 and 95.4\%~C.L. ($\Delta \chi^2<9.70$) allowed regions  
are shown in Fig.~\ref{fig:fluxnorm}. The plots are made versus $\beta P_H$ to make
the uncertainties in $R_{\rm SNO}^{CC}$ transparent. Once adiabatic constraints are included
for the MSW solutions with large mixing,  
the number of free parameters is three, and the best-fit line collapses to a 
best-fit point marked by the 
cross; we will discuss this in Section~\ref{sec:free}. 

For sterile neutrinos, $P_H$ is replaced by $\beta P_H$ in
Eqs.~(\ref{eq:RCl}), (\ref{eq:RGa}), and (\ref{eq:RSKs}). There is again
a family of solutions for the probabilities, but unlike the active case,
the values of $P_I$ and $P_L$ are fixed, and are the same as those in
the sterile solution with the $^8$B normalization fixed at unity,
Eq.~(\ref{eq:sterile}). Only the value of $P_H$ depends on $\beta$, with
$\beta P_H = 0.459$. Thus we see that in the sterile case the
intermediate energy neutrinos must be strongly suppressed regardless of
the value of the $^8$B normalization. This shows why the sterile
case requires the SMA solar solution, which has a strong suppression of
the intermediate energy neutrinos. Since $R_{\rm SNO}^{CC} = \beta P_H =0.459$, the
best fit solutions for sterile neutrinos give the same prediction for
$R_{\rm SNO}^{CC}$ for any value of the $^8$B flux normalization.

Because the SMA solution seems increasingly unlikely~\cite{s,oursno} 
for either active or sterile oscillations, we consider only active oscillations in 
the LMA and LOW solutions.

\section{MSW solutions with large mixing}

The consistency of particular neutrino oscillation solutions can be
tested using the probabilities $P_H$, $P_I$, and $P_L$ determined in the
previous sections. In MSW solutions the oscillation probability is~\cite{parke}
\begin{equation}
P(\nu_e \to \nu_e) = {1\over2} + ({1\over2} - P_c) \cos2\theta_M
\cos2\theta \,,
\label{eq:LZ}
\end{equation}
where $\theta_M$ is the effective mixing angle in matter at the creation
point of the electron neutrino and $P_c$ is the Landau-Zener probability 
for crossing from the
upper to the lower eigenstate as the neutrino propagates through the
matter in the sun. We consider only the LMA and LOW solutions, whose
 suppression of $^8$B neutrinos has very little energy dependence,
in agreement with the measured Super--K spectrum.

\subsection{Adiabatic solutions with $^8$B flux from SSM}

In a typical MSW solution with large mixing, all of the solar neutrinos
propagate adiabatically, which implies $P_c = 0$ in Eq.~(\ref{eq:LZ})~\cite{bethe}.
For neutrinos created in a region of the sun above the
critical density for a resonance to occur and that start far above resonance,
$\cos2\theta_M = -1$, which implies the oscillation probability is
\begin{equation}
P(\nu_e \to \nu_e) = {1\over2}(1 - \cos2\theta) = \sin^2\theta \,.
\label{eq:Pabove}
\end{equation}
For neutrinos that start well below resonance,
$\cos2\theta_M = \cos2\theta$ and
\begin{equation}
P(\nu_e \to \nu_e) =
{1\over2}(1 + \cos^22\theta) = 1 - {1\over2} \sin^22\theta \,.
\label{eq:Pbelow}
\end{equation}
For each part of the energy spectrum, we define $f_j$ ($j = H,I,L$) as
the fraction of those neutrinos that are created above resonance; then
\begin{equation}
P_j = f_j \sin^2\theta + (1-f_j)(1 - {1\over2} \sin^22\theta) \,.
\label{eq:prob}
\end{equation}
In Eq.~(\ref{eq:prob}) we have implicitly assumed that a negligible fraction of
neutrinos are created near the resonance. Although this is not strictly
true for a neutrino of a given energy, when averaged over the entire
spectrum Eq.~(\ref{eq:prob}) provides a good approximation to the overall
average probability $P_j$.

The condition for which neutrinos are created above resonance is
\begin{equation}
\delta m^2 \cos2\theta < 2 \sqrt2 G_F N_e E
= (1.52\times10^{-7} {\rm~eV}^2) N_e E \,,
\label{eq:res}
\end{equation}
where $N_e$ is the electron number density 
in the Sun in units of $N_A/$cm$^3$ and $E$ is the
neutrino energy in MeV; for a typical high or intermediate energy neutrino in the sun, 
$N_e \simeq 90$~\cite{SSM}. Thus, for a given $\delta m^2$ and $\theta$, Eq.~(\ref{eq:res})
defines the critical neutrino energy above which neutrinos are created
above resonance. 

The resonance condition depends directly on the neutrino energy.  For a
given $\delta m^2$ all of the neutrinos above a certain critical energy
will be created above resonance. Since neutrinos in one part of the
spectrum cannot be above resonance until all of the neutrinos with
higher energy are also above resonance, we see immediately that MSW
solutions must have $f_H \ge f_I \ge f_L$. Furthermore, we can define
three regimes: (i) some or all of the high energy neutrinos and none of
the low or intermediate energy neutrinos are created above resonance
($0 \le f_H \le 1$, $f_I = f_L = 0$), (ii) all of the high energy
neutrinos, some or all of the intermediate energy neutrinos, and none
of the low energy neutrinos are created above resonance ($f_H = 1$, $0
\le f_I \le 1$, $f_L = 0$), and (iii) all of the high and
intermediate energy neutrinos and some or all of the low energy
neutrinos are created above resonance ($f_H = f_I = 1$, $0 \le f_L \le
1$). It should be noted that although there are {\it a priori} three
fraction parameters ($f_H$, $f_I$, and $f_L$) in an MSW solution, $f_I$
cannot be nonzero unless $f_H = 1$ and $f_L$ cannot be nonzero unless
$f_I = 1$, so the $f_j$ are in fact equivalent to a one-parameter system with
each value of $\delta m^2$ corresponding to an unique set of $f_j$. 
By examining the expected differential rates
versus neutrino energy for a given part of the spectrum, the values of
the $f_j$ may be determined.
For example,
when $\sin^22\theta = 0.8$ ($\cos2\theta = 0.45$), $\delta m^2 =
10^{-4}$~eV$^2$ implies a critical energy of about 3~MeV, so that all of
the high energy neutrinos and none of the low and
intermediate energy neutrinos are created above resonance; this
corresponds to $f_H = 1$ and $f_I = f_L = 0$. Thus, $\theta$ and the
$f_j$ are effectively two free parameters. 

\begin{figure}
\centering\leavevmode
\mbox{\psfig{file=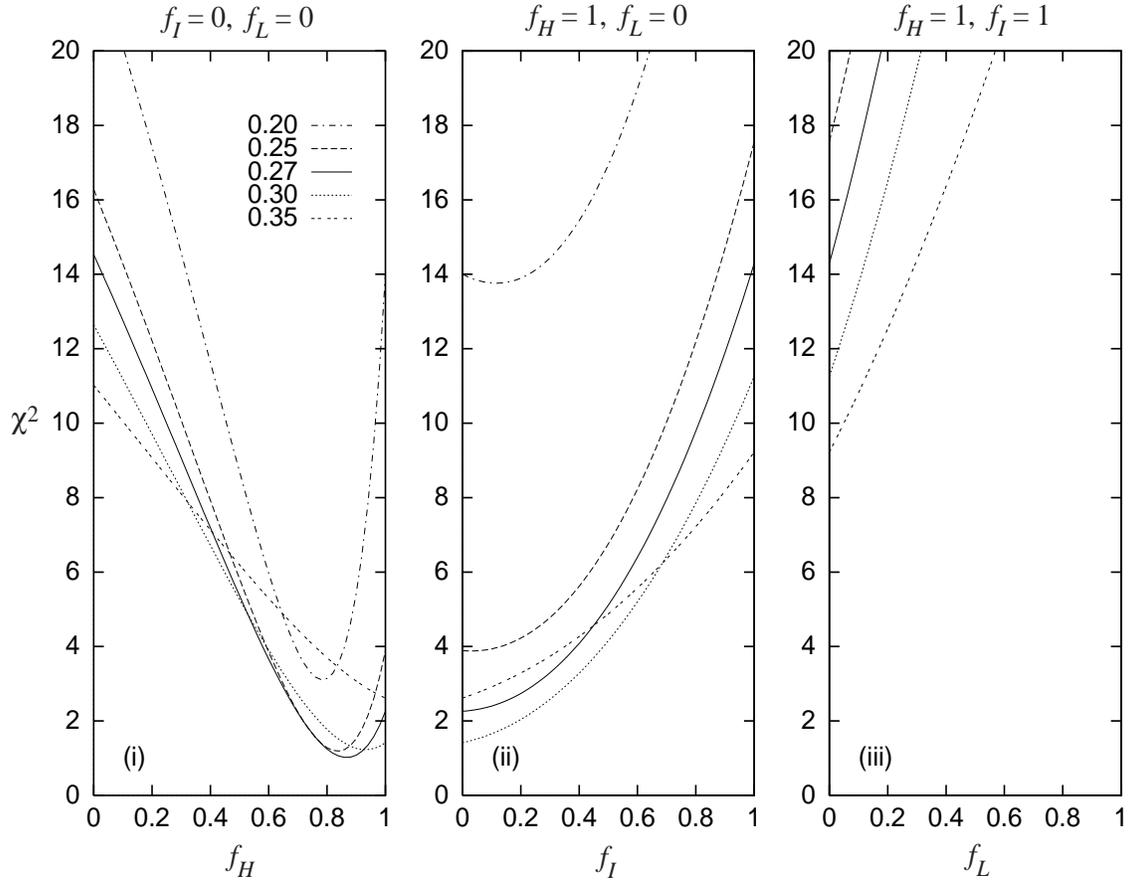,width=16cm,height=12cm}}
\medskip
\vspace{0.5cm}
\caption[]{$\chi^2$ for the LMA solution versus (i) $f_H$, with $f_I =
f_L = 0$, (ii) $f_I$, with $f_H = 1$ and $f_L = 0$, and (iii) $f_L$,
with $f_H = f_I = 1$, where $f_j$ is the fraction of $j$-type neutrinos
($j = H,I,$ or $L$) created above resonance, for various values of
$\sin^2\theta$. SSM fluxes are assumed ($\beta=1$).}
\label{fig:chi2}
\end{figure}

In Fig.~\ref{fig:chi2} we show $\chi^2$ versus the $f_j$ for these
three regimes for various values of $\sin^2\theta$, assuming the $^8$B
normalization is given by the SSM with the uncertainty in
Table~\ref{tab:spectrum}. The best fit is
\begin{equation}
\sin^2 \theta = 0.273\,,\  f_H = 0.874\,,\  f_I = f_L = 0 \,,
\label{eq:fit}
\end{equation}
which in terms of probabilities, Eq.~(\ref{eq:prob}) is
\begin{equation}
P_H =0.315 \,,\  P_I = 0.603 \,,\  P_L = 0.603\ \,.
\label{eq:fitp}
\end{equation}
The $\chi^2$/d.o.f. is 1.02/1, which corresponds to a goodness-of-fit
of 31\%. The associated oscillation amplitude is $\sin^22\theta = 0.79$, which is 
almost exactly the value obtained in the flux-constrained global analysis of solar neutrino
data in Ref.~\cite{analysis}.
For this value of $f_H$, 87\% of the
high energy neutrinos are created above resonance. An inspection of the
$^{37}$Cl and Super--K spectra shows that neutrinos with
energies above about 7~MeV are created above resonance, which by
Eq.~(\ref{eq:res}) gives $\delta m^2 \sim 2\times10^{-4}$~eV$^2$, a
value that corresponds to an LMA solution. However, from the flux-dependent
global analysis of Ref.~\cite{analysis}, this value of $\delta m^2$ is 
allowed only at the 99\% C. L. which indicates that the  
day and night spectra and the full resonance treatment with unaveraged
energy spectra are important in the global fit.

For the LOW solution, in our approximation all of the neutrinos are
created with $\cos2\theta_M = -1$ in Eq.~(\ref{eq:LZ}), so $f_H = f_I =
f_L = 1$ and $P_H = P_I = P_L = \sin^2\theta$. This condition is far
from the exact fit of Eq.~(\ref{eq:active}). The best fit for $P_H = P_I
= P_L$ with the SSM $^8$B flux normalization (i.e., $\beta =1$) is $P_j
= \sin^2\theta = 0.517$ with $\chi^2/$d.o.f. = $10.6/2$, which is excluded at
the 99.5\%~C.L. Thus the LOW solution is disfavored.

One might wonder why the values of $P_j$ in Eq.~(\ref{eq:fitp}) are not
the same as those in Eq.~(\ref{eq:active}) from the model-independent
analysis.  To understand this, we note that the
situations in which the intermediate and low energy neutrinos can have
different flux suppressions are cases (ii) and (iii) described above,
both of which require $f_H=1$. Working backwards from
Eqs.~(\ref{eq:active}) and (\ref{eq:prob}) with $f_H=1$ gives
$f_I=1.22$, which is unphysical.  Thus there is either an inconsistency
in the data from the different experiments or not all the SSM flux normalizations
 are correct.

\subsection{Adiabatic solutions with free $^8$B flux normalization}
\label{sec:free}

\begin{figure}[b]
\centering\leavevmode
\psfig{{file=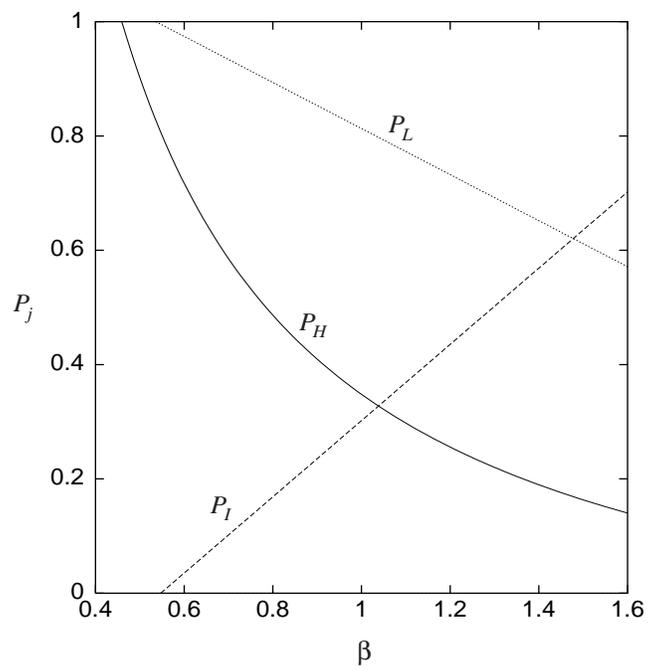,width=9cm,height=9cm}}
\medskip
\vspace{0.5cm}
\caption[]{Probabilities $P_H$, $P_I$, and $P_L$ for active neutrino oscillations
 that exactly reproduce the solar neutrino data, 
plotted versus $\beta$, the ratio of the $^8$B flux
normalization to its SSM value.}
\label{fig:beta}
\end{figure}

The MSW large angle solutions can also be tested when the $^8$B
normalization is allowed to vary, in which case the $P_j$ are given by
Fig.~\ref{fig:beta} as a function of the normalization factor $\beta$.
The figure shows that for $\beta \lsim 1$ ($^8$B flux less than the SSM
prediction) $P_H$ and $P_L$ are driven higher and $P_I$ lower than the
best fit for $\beta = 1$. This would be very hard to understand in the
LMA solution, because the high energy neutrinos typically start above
resonance (with $P = \sin^2\theta$) and the low and
intermediate energy neutrinos typically start below resonance (with $P =
1- {1\over2}\sin^22\theta$); since $\sin^2 \theta < 1 -
{1\over2}\sin^22\theta$ for $\theta < \pi/4$, this implies $P_H \le P_I
\le P_L$. On the other hand, for $\beta \gsim 1$, 
$P_I$ and $P_L$ are higher than $P_H$, which is qualitatively consistent with
the LMA solution. By imposing the condition $P_H \le P_I \le P_L$,
required for an MSW solution with $\theta < \pi/4$,
Fig.~\ref{fig:beta} shows that the LMA solution favors a $^8$B flux
normalization in the range $1.04 \le \beta \le 1.48$.

We note that for an MSW solution with $\theta > \pi/4$, all of the
neutrinos are created with $\cos2\theta_M = -1$ in Eq.~(\ref{eq:LZ}), so
in our approximation $f_H = f_I = f_L = 1$ and $P_H = P_I = P_L =
\sin^2\theta$. There is no value of $\beta$ for which this occurs in the
overall best fit (see Fig.~\ref{fig:beta}); hence this region of
parameter space is disfavored, and a normal mass hierarchy
($\theta<\pi/4\,,\delta m^2>0$) is selected. Similarly, as discussed
above, all three probabilities are equal in the LOW solution in our
approximation; the best fit for $\beta$ and $P_H = P_I = P_L$ free is
$\beta = 0.72$ and $P_j = \sin^2\theta = 0.540$ ($\theta>\pi/4$), with $\chi^2/$d.o.f. =
$9.16/1$. Thus the LOW solution is approximately 3$\sigma$ from the
overall best fit, similar to the findings of Ref.~\cite{analysis}.

In the above modelling with free $^8$B normalization, 
there are three free parameters for the large angle
solutions: the vacuum mixing angle $\theta$, the $^8$B normalization
$\beta$, and the fractions $f_j$ that determine which neutrinos start
above resonance. For oscillations to active neutrinos, we find that
the unique solution (with $\chi^2 = 0$) is 
\begin{equation}
\sin^2\theta = 0.206\,,\ \beta = 1.34\,,\ 
f_H = 1\,,\ f_I = 0.301\,,\ f_L = 0 \,.
\label{eq:fit2}
\end{equation}
This solution is very close to the values $\sin^2\theta = 0.21$
and $\beta = 1.31$ found in the global analysis of Ref.~\cite{analysis}, which also allowed
 the $^8$B flux be a free parameter. The $f_j$ of Eq.~(\ref{eq:fit2}) imply
that all of the high energy neutrinos and about 30\% of the
intermediate energy neutrinos are above resonance, which implies that the
critical energy lies on the $^7$Be line at 0.862~MeV; this translates into $\delta m^2 =
2\times10^{-5}$~eV$^2$ for $\sin^2\theta=0.206$ ($\sin^22\theta=0.654$),
 close to the best-fit value of Ref.~\cite{analysis}. Thus our simplified analysis
reproduces the results of a comprehensive global fit to the data. There is no exact
solution possible for oscillations to sterile neutrinos since that would
require $P_I \le P_H, P_L$, which is not allowed by the ordering
$P_H \le P_I \le P_L$; this explains why oscillations to sterile
neutrinos are disfavored for MSW solutions with large mixing angles.

\begin{figure}[b]
\centering\leavevmode
\psfig{{file=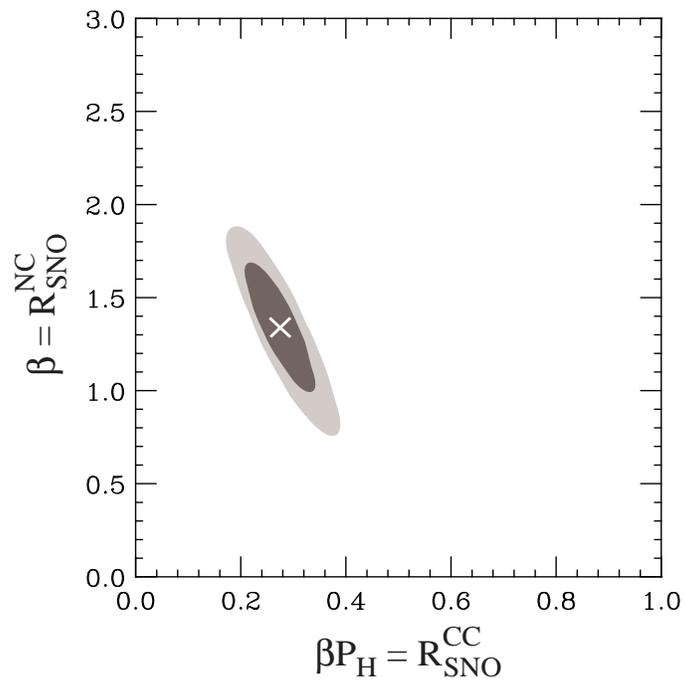,width=9cm,height=9cm}}
\medskip
\vspace{0.5cm}
\caption[]{The 95.4\% C. L. and 68.3\% C. L. allowed regions from an analysis with
$\beta$, $\theta$ and $f_j$ free. The best-fit point 
$(R_{\rm SNO}^{NC},R_{\rm SNO}^{CC})=(1.34, 0.28)$, 
is marked with a cross.}
\label{fig:snopred.ps}
\end{figure}

The probabilities corresponding to Eq.~(\ref{eq:fit2}) are (using~(\ref{eq:prob}))
\begin{equation}
P_H =0.206 \,,\  P_I = 0.532 \,,\  P_L = 0.673\ \,.
\end{equation}
The best-fit values of and uncertainties in 
$\beta$ ($R_{\rm SNO}^{NC}$) and $R_{\rm SNO}^{CC}$ are (see Fig.~\ref{fig:snopred.ps})
\begin{equation}
R_{\rm SNO}^{NC}=\beta=1.34\pm 0.34
\end{equation}
and
\begin{equation}
R_{\rm SNO}^{CC}=\beta P_H=0.28\pm 0.07\,.
\end{equation}
The three values of $P_j$, and $\beta$ from Eq.~(\ref{eq:fit2}) represent a unique 
best-fit in Fig.~\ref{fig:fluxnorm}, marked by a cross; the adiabatic constraints
 on the line of best-fit solutions  singles out one solution.

\section{Summary}

By parameterizing the expectations for the three types of solar
neutrino experiments ($^{37}$Cl, $^{71}$Ga and $\nu$--$e$ scattering),
in terms of three average survival probabilities for the high, intermediate, and 
low energy solar neutrinos, we have determined a unique best fit assuming
SSM fluxes. Accounting for the experimental and theoretical
uncertainties, allowed regions in the probability space were found. Our
analysis with the SSM fluxes yields a CC prediction for data/SSM for SNO of 
\begin{eqnarray}
R_{\rm SNO}^{CC} &=&0.35^{+0.16}_{-0.09}\ \ \ \ \rm{(active)} \\
            &=&0.46^{+0.13}_{-0.11}\ \ \ \ \rm{(sterile)}\,
\end{eqnarray}
where the uncertainties are $1\sigma$.
The prediction for oscillations to sterile neutrinos is flux-independent.
For some values of $R_{\rm SNO}^{CC}$ it could be possible to distinguish
between the active and sterile scenarios without using with the neutral-current
rate.

A scenario in which
the $^8$B flux is not affected by oscillations and is assumed to be
directly measured by Super--K is highly disfavored by the data for oscillations to
active neutrinos. 
Allowing the normalization of the $^8$B flux to be a free parameter, a
family of solutions was found that depend on the flux normalization
factor; $^8$B flux normalizations below the SSM imply that the
intermediate energy neutrino contribution must be more suppressed, while
normalizations above the SSM imply that the high energy neutrino
contributions are more suppressed.
The family of best-fit probabilities of Fig.~\ref{fig:beta} give a $\beta$-dependent 
prediction for the best-fit CC rate at SNO:
\begin{eqnarray}
R_{\rm SNO}^{CC} = \beta P_H &=& 1.21 R_{\rm SK}- 0.21 \beta \\
                        &=&  0.55 - 0.21 \beta\ \ \ \ \ \ \ \  (R_{\rm SK}=0.459)\,.
\end{eqnarray}
This equation can be inverted so that the central value of SNO CC measurement determines
 the central value of the NC rate, 
\begin{eqnarray}
R_{\rm SNO}^{NC}=\beta&=&5.85 R_{\rm SK}- 4.85 R_{\rm SNO}^{CC}\\
                      &=&2.69 - 4.85 R_{\rm SNO}^{CC}\ \ \ \ \ \ \ \  (R_{\rm SK}=0.459)\,.
\end{eqnarray}
Thus, it may be possible for SNO to obtain the $^8$B flux normalization without recourse to 
neutral-current measurements.

By imposing adiabatic constraints on our probability
parameterization with the $^8$B flux free, we found a unique solution
where all of the high energy and 30\% of the intermediate energy
neutrinos are created above resonance. This solution has a best-fit $^8$B flux
normalization \mbox{$\beta=1.34$}, and 
\begin{equation}
\delta m^2 = 2\times10^{-5}\  {\rm{eV}}^2\,,\  {\sin^22\theta} = 0.65\,,\  
\end{equation}
in the LMA region. 
These values are close to those found from more comprehensive global
fits to solar data~\cite{analysis}. The LOW solution and an inverted mass hierarchy
with $\theta<\pi/4$ and $\delta m^2<0$ are disfavored.
 From our best-fit, we predict
\begin{eqnarray}
\beta=R_{\rm SNO}^{NC}&=&1.34\pm 0.34\\
      R_{\rm SNO}^{CC}&=&0.28\pm 0.07\ \ \ \ \ \ \ \ ({\rm{active,\ best\ fit}}\ ^8{\rm{B\ flux}}).
\end{eqnarray}

\section*{Acknowledgments}

This research was supported in part by the U.S.~Department of Energy
under Grants No.~DE-FG02-94ER40817 and No.~DE-FG02-95ER40896, and in
part by the University of Wisconsin Research Committee with funds
granted by the Wisconsin Alumni Research Foundation.

\clearpage

\clearpage

\end{document}